%% file: ms.tex
\documentclass[12pt,preprint]{aastex}

\slugcomment {Accepted by ApJL}
\shorttitle{Spitzer Observations of CO$_2$ Ice}
\shortauthors{Bergin et al.}

\newcommand\etal{et al. }
\newcommand\be {\begin{equation}}
\newcommand\en{\end{equation}}

\input{macro.tex}

\slugcomment{Accepted by {\it The Astrophysical Journal Letters}}

\begin{document}

\title
{
Spitzer Observations of CO$_2$ Ice Towards Field Stars in the Taurus
Molecular Cloud
}

\author{Edwin A. Bergin\altaffilmark{1} 
Gary J. Melnick\altaffilmark{2},
Perry A. Gerakines\altaffilmark{3},
David A. Neufeld\altaffilmark{4},
Douglas C.B. Whittet\altaffilmark{5}
}

\altaffiltext{1}{University of Michigan, 825 Dennison Building, 501 E. University Ave.,
Ann Arbor, MI 48109-1090; email: ebergin@umich.edu}
\altaffiltext{2}{Harvard-Smithsonian Center for Astrophysics, 60 Garden St.,
Cambridge, MA 02138}
\altaffiltext{3}{
Astro- and Solar-System Physics Program, Department of Physics, University
of Alabama at Birmingham, 1300 University Blvd, CH 310, Birmingham, AL
35294-1170
}
\altaffiltext{4}{
Department of Physics and Astronomy, The Johns Hopkins University, 
3400 North Charles Street, Baltimore, MD 21218
}
\altaffiltext{5}{
Department of Physics, Applied Physics, and Astronomy, and New York Center
for Studies on the Origins of Life, Rensselaer Polytechnic Institute, Troy,
NY 12180
}

\begin{abstract}
We present the first {\em Spitzer} Infrared Spectrograph observations
of the 15.2 $\micron$ bending mode of CO$_2$ ice 
towards field stars behind a quiescent dark cloud.  CO$_2$ ice is detected towards
2 field stars (Elias 16, Elias 3) and a single protostar (HL Tau) with an
abundance of $\sim 15-20\%$ relative to water ice.  CO$_2$ ice is
not detected towards the source with lowest extinction 
in our sample, Tamura~17 (A$_V$ = 3.9$^m$).
A comparison of the Elias 16 spectrum with laboratory data demonstrates
that the majority of CO$_2$ ice is embedded in a polar
H$_2$O-rich ice component, with $\sim$15\% of CO$_2$ residing in an
apolar H$_2$O-poor mantle. This is the first detection of apolar CO$_2$ 
towards a field star.  
We find that the CO$_2$ extinction threshold is A$_V = 4^m \pm 1^m$,
comparable to the threshold for water ice, but significantly less than
the threshold for CO ice,
the likely precursor of CO$_2$.
Our results confirm CO$_2$ ice forms in tandem with H$_2$O ice
along quiescent lines of sight.
This argues for CO$_2$ ice formation via a mechanism similar to
that responsible for H$_2$O ice formation, viz. 
simple catalytic reactions on grain surfaces.
\end{abstract}


\section{Introduction}
\label{sec_intro}

%One of the early highlights from the {\em Infrared Space Observatory} ({\em ISO})
%was the detection of \COtwo\ ice in the interstellar medium (ISM) 
%\citep{degraauw_co2}.
Observations by the {\em Infrared Space Observatory} ({\em ISO})
demonstrated that \COtwo\ ice is a ubiquitous
component of the interstellar medium (ISM), 
with typical abundance of $\sim$15 -- 25\% relative
to \HtwoO , the dominant ice component \citep{gerakines_co2, nummelin_co2}. 
However, the origin of this common grain mantle constituent
remains uncertain.  Gas-phase production,
with subsequent freeze-out, is believed incapable of reproducing the observed 
abundance.  In the laboratory CO$_2$ forms quite 
readily via ultraviolet (UV) photolysis or ion bombardment of astrophysical
ice mixtures; along with the thermal processing of ices, 
these mechanisms have been suggested as
possible formation routes (d'Hendecourt et al. 1986; Sandford et al. 1988;
Palumbo et al. 1998).
Even if UV photons and cosmic rays are not present with sufficient fluxes, an
alternative mechanism is grain surface chemistry
(Roser et al. 2001; Frost, Sharkey, \& Smith 1991, and references therein).  
Observations of grain
mantles indicate that there is competition 
between hydrogenation and oxidation of atoms on grain surfaces
\citep[][and references therein]{ewine_iso}. 
Nowhere
is this more clear than for carbon-bearing molecules.  \COtwo\ ice is less
volatile than CO \citep{sandford_eb} and its 
formation locks
carbon in saturated form on grains, 
stopping any subsequent hydrogenation 
towards more complex molecules on grains or in the gas.  
Thus, an understanding of the formation
mechanisms of \COtwo\ ice is needed in order to understand the limits of
grain surface chemistry in producing more complex organics.

One way to constrain the origin of CO$_2$ ice in the ISM is to use bright
field stars located behind molecular cloud material as candles that probe
material remote from embedded sources,
where, at suitably high extinction, ices are
unlikely to be exposed to significant UV radiation or 
heating.
%Observations of field stars in the mid-infrared ($3-20 \mu$m), where the
%primary vibration modes of astrophysical ices lie, require greater
%sensitivity compared to those of embedded protostars,
%which have excess radiation above the
%stellar photosphere in the infrared. 
% As such ISO was only capable of
{\em ISO} 
detected the CO$_2$ 4.27 $\mu$m $\nu_3$ stretching mode towards only two 
K giants, Elias 13 and 16 \citep{whittet_co2, nummelin_co2}.
However, the detection of CO$_2$ ice towards any field star  
demonstrated that radiative processing of ices is unlikely to be responsible
for the CO$_2$ ice production in quiescent material 
\citep{whittet_co2}.  

In this paper we report, and discuss the implications of,
the observation of the $\nu_2 $ bending mode  of CO$_2$ ice
at 15.2 $\mu$m
towards 3 field stars and 1 protostar in the Taurus molecular cloud using
NASA's {\em Spitzer Space Telescope}.  
%The observations are 
%described in \S2, and
%in \S 3 we discuss the amount of \COtwo\ embedded within
%polar and apolar ice components.  In \S 4 we examine 
%the correlation of integrated opacity with 
%visual extinction, which is another 
%constraint on the formation mechanism.  In \S 5 we discuss
%the implications of our result for the formation of CO$_2$ ice.

\section{Observations and Results}

Observations of each source (see Table~1) were obtained in 2004 using the
{\em Spitzer} Infrared Spectrograph, hereafter IRS \citep{houck_irs}.
%Tamura 17 was observed on Feb. 27, Elias 16 and Elias 3 on March 3, 
%and HL Tau on Oct. 10.  
Each object was observed in staring mode
in the short-wavelength high-resolution mode spectrograph
(Short-Hi;SH) which has coverage from 10 to 19 $\mu$m with $\lambda/\Delta 
\lambda \sim$ 600.  
%The normal observing mode includes 2 separate nodding
%observations with the telescope offset by 1/3 of the slit length
%in the spatial direction.  
Each exposure was taken using a per
cycle integration time of 120 seconds  with 4 cycles.
All data were reduced using the IRS Team's SMART program
\citep{higdon_smart}, starting with data products from pipeline version
13.  Our data reduction process mirrored that described
by \citet{watson_irs} (see \S 2 in that paper).
%Fringes were not removed from the data.  

Figure~1 shows the IRS spectra with 
a clear CO$_2$ absorption feature at
$\sim$15 $\mu$m for Elias~16, Elias~3, and HL~Tau.
For Tamura 17, the star tracing the lowest extinction in our
sample (Table~1), the stellar continuum
is detected, but no absorption feature is observed. 
%H$_2$O and CO ice has previously been detected towards 
%all field stars\citep{teixeira_h2o},
%but CO ice (e.g. CO 
%has not been detected towards HL Tau).
For CO$_2$ the 15 $\mu$m feature optical depth is derived 
by fitting a multi-order polynomial to determine the continuum. In 
general polynomials with orders 2--4 (higher order for Elias~16) 
were required to account
for the mismatch in the continuum on the longward side of the CO$_2$ line
caused by wings of the broad 18.5 $\mu$m silicate feature.  
Column densities are estimated using N(CO$_2$) = 
$\int \tau$($\tilde{\nu}$ ) d$\tilde{\nu}$/$A_{15.2\mu m}$, where
$\tilde{\nu}$ is the frequency in wavenumber units and 
a value of $10^{-17}$ cm/molecule is assumed for
$A_{15.2 \mu m}$ \citep{gerakines_co2str}.  
The column density 
derived towards Elias~16 is identical, within errors, to that derived
earlier by Whittet et al. (1998) using the {\em ISO} detection of the
4.27 $\mu$m line.

\section{Elias 16 Line Profile}

The Elias 16 spectrum possesses a sufficient signal to noise 
ratio to permit an
analysis of the ice composition 
\citep{gerakines_co2, nummelin_co2}.  
The 4.27 $\mu$m feature detected by {\em ISO} constrained the
majority of CO$_2$ to lie within a single polar ice component
with H$_2$O:CO$_2$:CO (100:20:3) at 20 K (Whittet et al. 1998).
This contrasts with
CO ice, which resides primarily in an apolar component
(Chiar et al. 1995). 

The \COtwo\ bending mode
is more sensitive than the stretching mode to the ice composition 
\citep{ehren_co2}, and the spectrum, shown in Figure~2, exhibits clear 
asymmetry.  We have used a similar set of laboratory interstellar
ice analogs (Ehrenfreund et al. 1996, 1997) 
with a $\chi^2$ minimization routine
(Gerakines et al 1999) to constrain the ice composition.
Because the gas temperature of the Taurus cloud
is $\sim$10~K we only present fits using ice analogs at 10 K and
%Because the temperature of molecular gas in Taurus
%is typically $\sim$10 K, we have only used laboratory data with
have excluded any spectra that include 
appreciable amounts of O$_2$. Molecular oxygen has yet to be
detected in the ISM either by direct detection in the gas-phase or
by indirect methods on grain surfaces
\citep{vanden_o2, goldsmith_o2, pagani_o2}.

Our best fit (Fig. 2) requires two components, a broad feature 
consistent with CO$_2$ ice within the polar water mantle
(H$_2$O:CO$_2$ 100:14) and
a narrow component of CO$_2$ embedded in an apolar matrix
(CO:CO$_2$ 100:26). 
The majority of CO$_2$ resides in the polar component, while
apolar CO$_2$ ice accounts for $\sim$15\% of 
the CO$_2$ ice column.\footnote{This apolar component is effectively 
hidden within the {\em ISO} 4.27 $\micron$ spectrum of Elias 16 
as there is little difference in the fits obtained with a purely
polar mixture \citep{whittet_co2} and those including a weak
apolar component such as that found here.}
%Thus CO$_2$ appears to be a minor constituent
%in both the polar and apolar mantle, but the majority of solid CO$_2$ is embedded
%with water ice.  This is the first evidence for apolar CO$_2$ ice
%towards a field star.

\section{Ice Extinction Threshold}

It is known from previous studies of ice features in Taurus that the absorption strength
correlates with extinction. The correlation line intercepts the extinction axis at a
positive value,  i.e. there exists a threshold extinction below which the ice feature is
not seen, presumably because the grains in the more diffuse outer layers of the cloud are
not mantled (e.g. Chiar \etal\ 1995, Whittet \etal\ 2001 and  references therein). 
Figure 3 compares plots of column density vs. extinction for CO$_2$ and  CO. In the case
of CO$_2$, we combine  both {\em Spitzer} and {\em ISO} observations.
%bending-mode results from the present work with those for the
%stretching mode from ISO SWS observations. 
The field-star data suggest a
correlation yielding a threshold extinction $A_V = 4^m\pm1^m$, i.e. not significantly
different from the value of $3.2^m\pm0.1^m$ reported for water-ice (Whittet \etal\ 2001). In
contrast, the threshold estimated for CO ($6.8^m\pm1.6^m$) appears to be significantly
larger. These results are consistent with a model in  which most of the CO$_2$ is in the
polar H$_2$O-rich component, whereas most of the CO is in the apolar,
H$_2$O-poor component. A larger threshold is expected for the latter because of its
greater volatility, requiring a greater degree of screening from the external radiation
field.\footnote{Note that the pre-main-sequence star HL Tauri does not follow the field-star trend
in either frame of Fig.3. Much of the extinction toward this object evidently arises in a
circumstellar disk (e.g. Close \etal\ 1997). Temperatures in the disk likely range from
$\sim 100$K, where CO is entirely in the gas phase (see Gibb \etal\
2004a) to much higher
temperatures where all ice mantles are sublimed. That CO$_2$ is detectable in solid form
toward HL Tau is consistent with its residence in a polar matrix.}

Figure 4 plots ice-phase column densities for CO$_2$ vs. H$_2$O and CO. 
%again combining
%results from SST and ISO SWS. 
In the case of CO$_2$ vs. H$_2$O, there is
a general trend -- linear least-squares fits to field stars and YSOs
are similar and intercept close to the origin. In contrast, the CO$_2$ vs. CO plot shows
a tendency to divide into two distinct trends (Gerakines \etal\ 1999).
For a given CO ice column, field stars have a lower CO$_2$ ice column than
massive YSOs. Grains in front of field stars are covered by polar and apolar mantles.
In contrast,  in the warm envelopes of 
massive YSOs, the dominant factor is likely to be
sublimation of apolar CO-rich ices
(although some CO$_2$ might also be produced by energetic processes). The lower-mass YSOs
show an intermediate distribution.

\section{Implications for CO$_2$ Ice Formation}

The {\em Spitzer} data on field stars revealed that
(1) most of the CO$_2$ ice is embedded within the water
ice mantle and (2) the CO$_2$ extinction threshold is closer to the
threshold for water
ice than that of its presumed precursor molecule, CO.  These two results
strongly suggest that CO$_2$ ice formation occurs in tandem with
that of water ice. Water ice is believed to form via surface reactions
during phases when gas is rich in atomic hydrogen
and atomic oxygen.  Observations of water vapor imply 
that atomic oxygen must be depleted in dense evolved well
shielded molecular regions; the low abundance of 
gas-phase water inferred by SWAS (Snell et al. 2000)
and ODIN (Olofsson et al. 2003) can only be accounted for in
models where nearly all available gas-phase atomic oxygen becomes
locked on grains (Bergin et al. 2000).  
%If
%a substantial  atomic oxygen component is present in dense
%gas then simple and
%well known gas phase reactions would efficiently create water with
%an abundance orders of magnitude higher than observed. 

Thus H$_2$O and CO$_2$ ice formation must occur during
the early lower density formative stages of the cloud.
Recent models of molecular cloud formation behind
shock waves by \citet{bergin_cform} 
may therefore be useful in setting constraints on ice formation.
They found that H$_2$ formation occurs 
at earlier times than gas-phase CO formation because
H$_2$ efficiently self-shields while
CO formation requires dust shielding 
(A$_V \sim 0.7$ mag). 
At such low cloud depths ice mantle formation would
be retarded by UV photodesorption.
However, for A$_V$ $> 1.0^m$ the effects of photodesorption are 
greatly reduced.  In this scenario 
CO gas-phase formation precedes both CO and water ice mantle formation. 
This qualitatively answers several questions.
Because most available gas-phase carbon would be locked in gas-phase CO, 
it would preclude a high abundance of methane ice,
in accord with observations \citep{gibb_iso}.   It also
allows for H, O, and CO to be present on the grain surface to
react via simple catalytic reactions to create H$_2$O and 
CO$_2$.\footnote{CH$_3$OH ice is not detected towards field stars \citep{chiar_ch3oh}.
Thus observations would suggest that grain surface formation of
CH$_3$OH is inefficient 
under low temperature quiescent conditions.
Laboratory experiments investigating the hydrogenation
sequence for CO, CO $\rightarrow$ HCO $\rightarrow$ H$_2$CO 
$\rightarrow$ CH$_3$OH (excluding some intermediate products), 
are discussed by \citet[][and references therein]{hiraoka_ch3oh}.
} 

One key question remains: how to account for the presence
of apolar CO$_2$ ice?  If CO$_2$ forms in tandem with H$_2$O in oxygen
rich gas then how
is a separate component of CO$_2$ formed with little H$_2$O?
This implies the presence of \ion{O}{1} in gas with little \ion{H}{1}. 
There are at least 2 scenarios that could account for 
this mantle structure: line of sight structure in the abundance of
atomic hydrogen or atomic oxygen (or perhaps both).
To examine the question of line of sight structure in atomic hydrogen,
in Elias~16 the abundance of
CO$_2$ in the apolar mantle is $\sim 3 \times 10^{-6}$ (all abundances
relative to
H$_2$): thus gas-phase \ion{H}{1} would need to
fall below this value to stop \ion{O}{1} hydrogenation.
Atomic hydrogen is expected to have near constant space density 
in molecular clouds, $n_{\rm{H\,I}} \sim 1 - 5$ 
cm$^{-3}$ (see Goldsmith \& Li 2005).  Thus, 
the \ion{H}{1} abundance should inversely follow density variations
along the Elias~16 line of sight.
However, 
the density would need to be $> 10^{6}$ cm$^{-3}$ for the
abundance of \ion{H}{1} to
fall below that required for \ion{O}{1}. This density
is characteristic of a condensed molecular core, which is not detected
towards this line of sight \citep{cerni_taurus}, and is therefore 
implausibly high.  
Thus \ion{H}{1} abundance structure 
is insufficient to account for apolar CO$_2$.

Line of sight structure in the \ion{O}{1} abundance provides a more
plausible solution to this issue. 
If atomic oxygen were absent in the densest regions with high
extinction, by water ice formation and other additional
solid-state reservoirs,
then oxygen hydrogenation could have halted in these regions. 
Oxidation could continue in layers with lower extinction and density.
%is that atomic oxygen can be absent
%in the densest regions with high extinction, but  could still be present with 
%an abundance comparable to or greater than \ion{H}{1} at 
%low to moderate A$_V$.  
For instance, assuming a
density of 10$^4$ cm$^{-3}$ for the Elias 16 line of sight \citep{bergin95},  
the atomic oxygen abundance would need to be $\gtrsim 10^{-4}$ 
to be higher than \ion{H}{1}, which is conceivable given the available \ion{O}{1} 
\citep[$\sim 8 \times 10^{-4}$;][]{jensen_oxy}.  
Thus CO oxidation could continue in outer
layers rich in atomic oxygen. 
There is some evidence for large \ion{O}{1} columns towards molecular
clouds that may trace low density layers \citep[][and references
therein]{caux_o, lis_o, li_o}.
This qualitative model can be tested by future  
higher signal-to-noise {\em Spitzer} observations 
of low extinction field stars (e.g. Elias~3).
For instance, line of sight structure in the oxygen abundance would
predict the existence of apolar CO$_2$ at moderate optical depths, 
even at those below the CO ice threshold.
Alternately, under the assumption that lower A$_V$ implies 
lower density, then decreasing amounts of apolar CO$_2$ would
suggest a relation to the declining abundance of atomic hydrogen with
increasing density.

\acknowledgements

We are grateful to the referee for useful comments.
DW acknowledges financial support from the NASA NSCORT program (grant
NAG5-12750).
D.A.N. gratefully acknowledges the support of a grant from NASA's LTSA
program
This work is based on observations made with the 
Spitzer Space Telescope, which is operated by the Jet Propulsion
Laboratory, California Institute of Technology, under NASA contract 1407. 
We are grateful to D. Watson and J. Green for aid in the data reduction and
are thankful for support from G. Fazio and the IRAC team.
These observations
were performed as part of the IRAC Guaranteed Time Observations
program.

\begin{figure}
\plotone{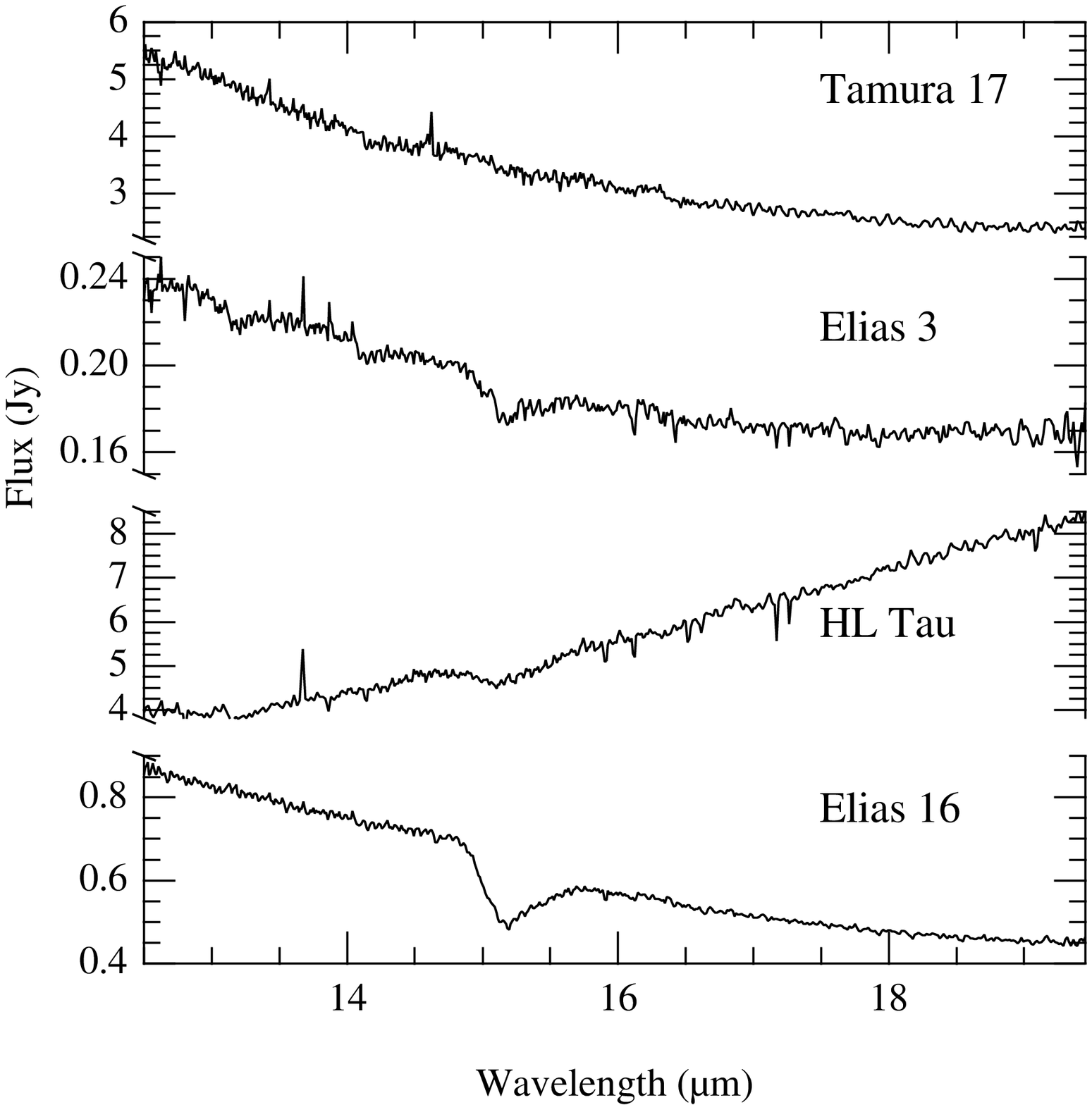}
\caption{{\em Spitzer} IRS Short-HI spectra towards background
stars Tamura 17, Elias~3, and Elias~16; HL~Tau is a Class I pre-main
sequence star.  These spectra are shown without any baseline
subtraction.}
\end{figure}

\begin{figure}
\plotone{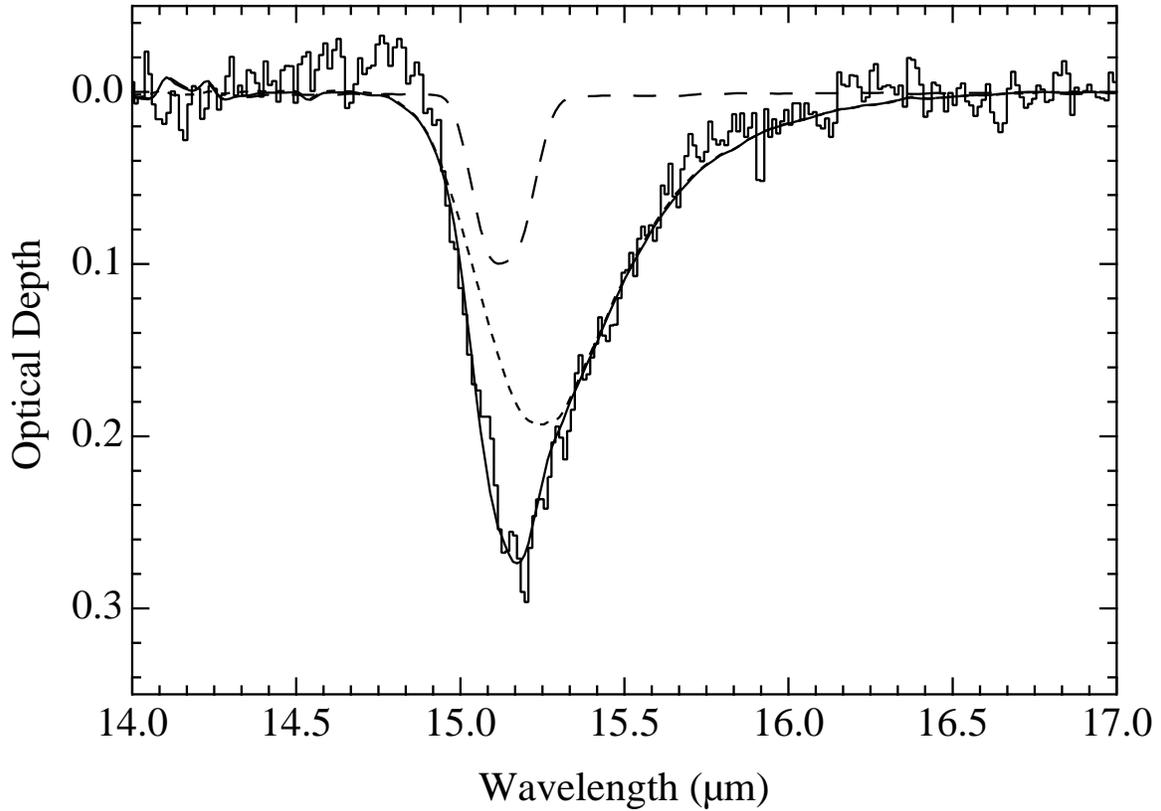}
\caption{Optical depth of CO$_2$ ice absorption seen towards Elias~16
(histogram).  Best fit line profile using laboratory analogs of
interstellar ice mixtures at 10 K (solid line).
The best fit requires two components: CO$_2$ embedded in a water rich
(polar) mantle, shown as the short-dashed line (laboratory ice analog mixture
H$_2$O $+$ CO$_2$ 100:14).  The second component, shown as long-dashed
line is CO$_2$ embedded in an apolar mantle (laboratory ice analog mixture
CO $+$ CO$_2$ 100:26).  
}
\end{figure}

\begin{figure}
\epsscale{0.6}
\plotone{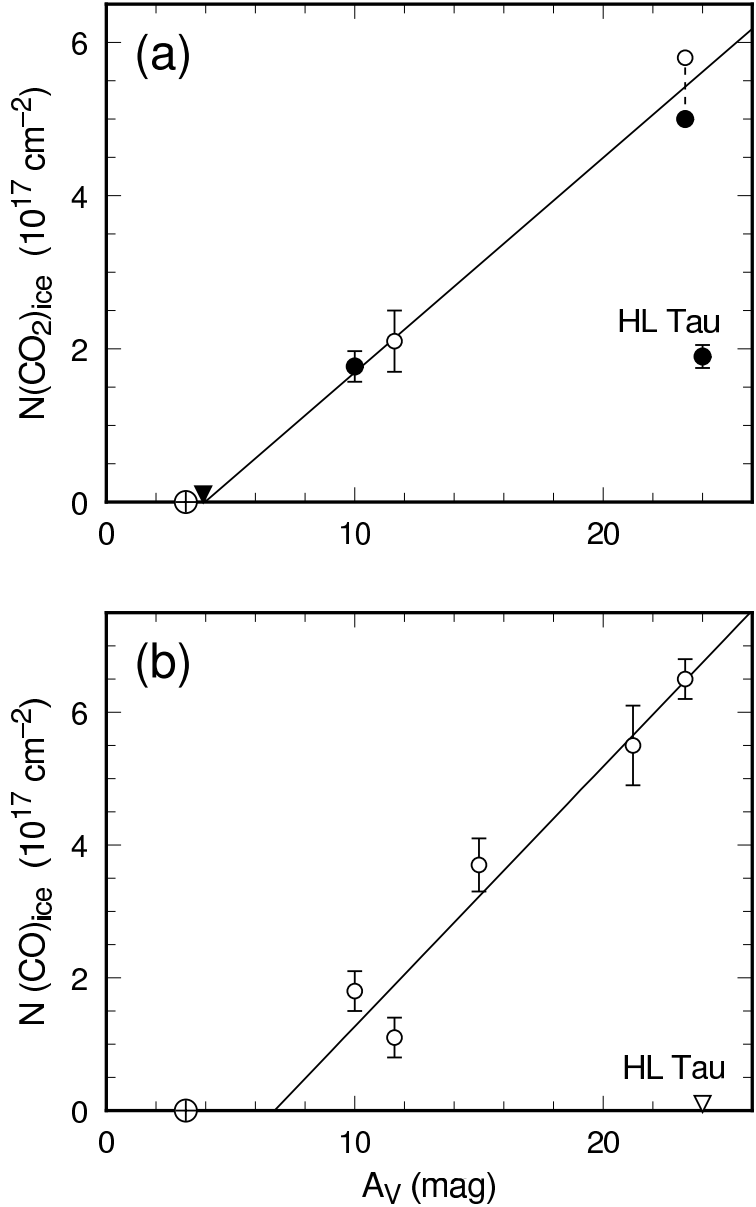}
\caption{
Plots of ice column density vs. visual extinction for (a) CO$_2$ and (b) CO.
Open and filled symbols in (a) represent stretching and bending vibrational
modes of CO$_2$, observed with the ISO SWS (Gerakines \etal\ 1999; Nummelin
\etal\ 2001) and SST IRS (this paper), respectively. The points for one
object (Elias~16) observed in both vibrational modes are joined by a
vertical dashed line. Triangles indicate upper limits. 
The circled
cross on the $A_V$ axis indicates the locus of the threshold extinction for
H$_2$O-ice ($A_V = 3.2$; Whittet \etal\ 2001). The diagonal line in each
frame is the linear least-squares fit to field stars.
}
\end{figure}

\begin{figure}
\epsscale{0.75}
\plotone{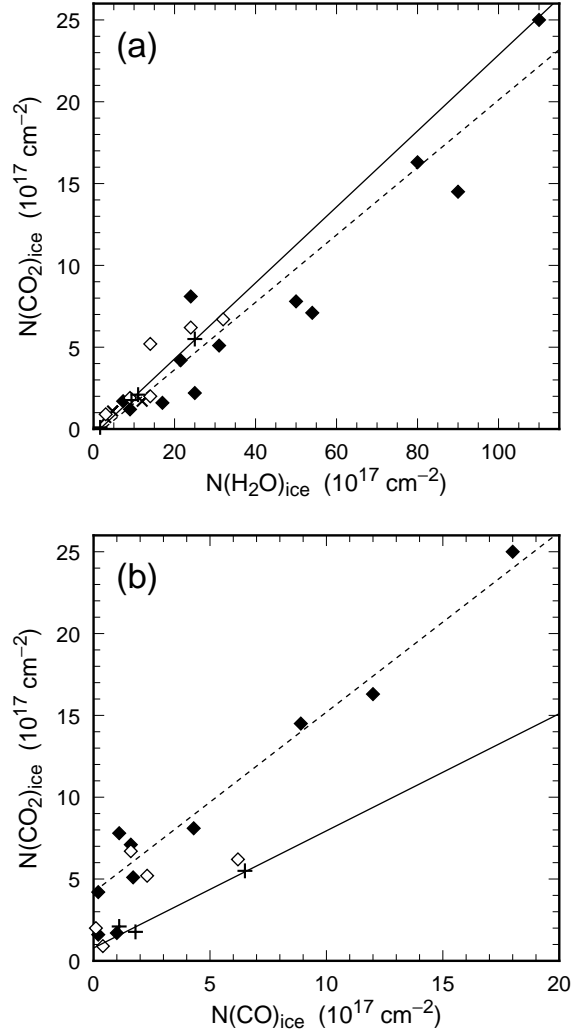}
\caption{
Plots of $N$(CO$_2$) vs.(a) $N$(H$_2$O) and (b) $N$(CO) for all sources with
data available from SST (this paper) or ISO SWS (Gerakines \etal\ 1999;
Nummelin \etal\ 2001). The symbols have the following meanings: Taurus field
stars ($+$); Galactic Center field stars ($\times$); low-mass YSOs (open
diamonds); high-mass YSOs (solid diamonds). The solid diagonal line in each
frame is the linear least-squares fit to field stars. The dashed line in (a)
is the fit to YSOs only (both high and low mass). The dashed line in (b) is
the relation  $N({\rm CO_2}) = 4.2 \times 10^{17} + 1.1N({\rm CO})$ proposed
by Gerakines \etal\ (1999) for high-mass YSOs.
}
\end{figure}

\begin{deluxetable}{lrrrrrr}
\label{tab1}
\tablecolumns{7}
\tabletypesize{\footnotesize}
\tablewidth{5.0in}
\tablecaption{Observation Parameters and Results}
\tablehead{
\colhead{Source} &
\colhead{Obs. Date} &
\colhead{$\int \tau(\tilde{\nu}) \Delta \tilde{\nu}$\tablenotemark{a}} &
\colhead{N(CO$_2$)\tablenotemark{b}} &
\colhead{$\rm{\frac{N(CO_2)}{N(H_2O)}}$} &
\colhead{$\rm{\frac{N(CO_2)}{N(CO)}}$} &
\colhead{A$_V$} 
}
\startdata
Elias 16\tablenotemark{f} & Mar. 3, 2004 & 5.0$\pm$0.1  & 5.0 & 0.21 & 0.8 & 23.3$^m$ \\
Elias 3\tablenotemark{f} &Mar. 3, 2004  & 1.8$\pm$0.1  & 1.8 & 0.20 & 1.0 & 10.0$^m$   \\
Tamura 17\tablenotemark{f} & Feb. 27, 2004 & $<$0.1(3$\sigma$) & $<$0.12 & $<$0.10 & \nodata & 3.9$^m$  \\
HL Tau & Oct. 10, 2004 &1.9$\pm$0.1 & 1.9 & 0.14 & \nodata & 24.0$^m$ 
\enddata
\tablenotetext{a}{
For Elias~16, Elias~3, and HL Tau integrated optical 
depths are calculated by a direct integration over the profile.
For Tamura 17, we have estimated the opacity limit by fitting a series of
Gaussians with fixed width and line center determined by the Elias~3
feature.  The absorption depth is variable and the
minimum optical depth that fits the noise is used to 
estimate the 3$\sigma$ integrated opacity. 
Units are cm$^{-1}$.
}
\tablenotetext{b}{In units of 10$^{17}$ cm$^{-2}$.}
\tablenotetext{f}{Denotes field star.}
\tablecomments{
$N$(CO) data for
field stars and the limiting value for HL~Tau are from Chiar \etal\ (1995)
and Tegler \etal\ (1995), respectively. The $A_V$ value for HL~Tau is from
Close \etal\ (1997), those for field stars are calculated from the $J-K$
color excess assuming $A_V = 5.3E_{J-K}$ (Whittet \etal\ 2001).
}
\end{deluxetable}

\end{document}

%% file: macro.tex
\def\m17{M~17}               % M 17    
\def\HtwoO{H$_2$O}             % H2O 
\def\COtwo{CO$_2$}       % C02
\def\m{\ts {\rm m}}

\def\swash2o{$1_{10} - 1_{01}$}             

\let\ts=\thinspace